\pdfoutput=1
\documentclass[conference]{IEEEtran}\IEEEoverridecommandlockouts

\usepackage{amsmath,amssymb,amsfonts}
\usepackage{soul}
\usepackage{graphicx}
\graphicspath{{diagrams/}}
\usepackage{cite}
\usepackage{url}
\usepackage{booktabs}
\usepackage{flushend}
\usepackage[utf8]{inputenc}
\usepackage[T1]{fontenc}
\usepackage{gensymb}
\usepackage{enumitem}

\usepackage{listings}
\usepackage{caption}
\lstdefinelanguage{json}{
  morestring=[b]",
  showstringspaces=false,
}
\setlist[itemize]{leftmargin=*}

\usepackage[letterpaper, top=19.1mm, bottom=19.1mm, left=19.1mm, right=19.1mm]{geometry}  
\usepackage{afterpage}  

\let\oldtexttt\texttt
\renewcommand{\texttt}[1]{{\small\oldtexttt{#1}}}

\begin{document}
\newgeometry{top=25.4mm, bottom=19.1mm, left=19.1mm, right=19.1mm}

\title{Using Large Language Models for Black-Box Testing of FMU-Based Simulations
}

\author{%
\IEEEauthorblockN{%
Abdullah Mughees\IEEEauthorrefmark{1},
Gaadha Sudheerbabu\IEEEauthorrefmark{1},
Tanwir Ahmad\IEEEauthorrefmark{1},\\
Dragos Truscan\IEEEauthorrefmark{1},
Mikael Manng{\aa}rd\IEEEauthorrefmark{2},
Kristian Klemets\IEEEauthorrefmark{3}}
\IEEEauthorblockA{\IEEEauthorrefmark{1}\AA bo Akademi University, Finland
\texttt{(firstname.lastname@abo.fi)}}
\IEEEauthorblockA{\IEEEauthorrefmark{2}Novia University of Applied Sciences, Finland 
\texttt{(firstname.lastname@novia.fi)}}
\IEEEauthorblockA{\IEEEauthorrefmark{3}University of Turku, Finland
\texttt{(firstname.lastname@utu.fi)}}
}

\maketitle
\begin{center}
\small
Accepted for presentation at the 2026 European Control Conference (ECC),
July 7--10, 2026, Reykjavík, Iceland.
The final version will appear in the ECC Proceedings.
\end{center}

\IEEEoverridecommandlockouts

\begin{abstract}
We propose a human in the loop approach for black-box testing of Functional Mock-up Units (FMUs) using Large Language Models (LLMs). The goal is to reduce the manual effort in defining test scenarios for dynamic simulation models and to improve the interpretability of results. The approach takes the functional and interface specifications of an FMU as input, and prompts an LLM to generate structured \emph{scenario goals} in Given-When-Then format that define the initial input conditions of the simulation, a possible change in those conditions, and the expected output behaviour of the system against those changes. The corresponding \emph{scenario plans} specify input patterns and add assertion oracles that describe expected output patterns defined in scenario goals. The approach generates a complete input time series for the scenario plans, runs the FMU simulation, and evaluates assertions on the recorded outputs. It produces human-readable logs and plots that show statistics for each scenario with overlays, aggregate pass rates, and per-goal outcomes. The generated scenarios and results are stored for evaluation and later re-execution. We evaluate the approach on a Lube Oil Cooling system and discuss design choices that make the approach practical for everyday use. Results suggest that LLM-assisted scenario generation can facilitate automatic test design and verification of dynamic simulation models.
\end{abstract}


\section{Introduction}
Modern engineering systems 
increasingly rely on model-based design, where components are developed and validated through simulation before physical implementation. In such workflows, different teams or organizations often use heterogeneous modeling and simulation tools, creating a need for standardized model exchange and co-simulation.
The Functional Mock-up Interface (FMI) \cite{FMI} addresses this challenge by defining a tool-independent standard for exchanging and executing simulation models as Functional Mock-up Units (FMUs). An FMU is a ZIP archive that typically contains: (i) model binaries, (ii) model description (specifying variable names, data types, causality, variability, start, stop, step values), and (iii) optional resources (such as parameter files, documentation). 
While this encapsulation enables flexible model exchange, it also introduces significant challenges for verification and validation of FMUs.
In particular, the internal equations and solver details of an FMU are not accessible, which makes traditional white-box testing infeasible.
Therefore, testing relies on the functional specifications and interface description.

Recent developments in Large Language Models (LLMs) have shown that they are well-suited to extract and structure information from textual and XML inputs.
In this paper, we propose a black-box testing approach that uses an LLM to: (i) extract I/O constraints and important data from FMU metadata and specification documents; and (ii) generate a diverse set of scenario goals and scenario plans. To ensure consistency and correctness of the generated output, we adopt a human-in-the-loop approach; a domain expert reviews generated goals and plans, and once approved, scenario plans are instantiated into complete input time-series scenarios, executed against the FMU, and evaluated based on assertions provided by the LLM. We evaluate the approach with an example and discuss design choices that make the method practical for everyday use. To that extent, we define an approach for evaluating the quality of the generated scenarios based on a custom mutation analysis method which we consider as a second contribution of this work.

The paper is structured as follows: Section~\ref{sec:related-work} reviews the related work that has been done in the field. Section~\ref{sec:preliminaries} provides a detailed background on the LLM related concepts and introduces a running case study. Section~\ref{sec:approach} presents our approach and the implementation. Section~\ref{sec:validation} evaluates the approach  and analyzes the results. Lastly, Section~\ref{sec:conclusions} discusses limitations of the applied approach and future work directions. 

\section{Related work}
\label{sec:related-work}
Research on automatic test generation and prioritization for Simulink models shows that search-based strategies for dynamic system behaviours can improve coverage and fault detection when model artefacts are available \cite{matinnejad2018test}. These methods assume that there is complete access to Simulink model internals and structural heuristics. However, in the case of FMUs, the internal working of the system is usually hidden, and the only source of information is the model description file and functional specification documents.

Recent work on intention-guided LLM unit tests \cite{nan2025test} shows that test intention improves the relevance and accuracy of generated tests in code-centric settings. 
Rangeet et al., \cite{pan2025aster} present the ASTER tool that supports automated generation of unit tests with LLMs guided by static program analysis. Meta's TestGen-LLM \cite{alshahwan2024automated} tool utilises LLMs to enhance existing human-written tests by generating additional test cases, thereby increasing test coverage through an approach known as 'Assured LLM-based Software Engineering' (Assured LLMSE). In their approach, human review is applied only for the test cases that the tool can guarantee to improve on the existing code base. Compared to these studies, our workflow offers a stepwise approach for black-box testing of FMU models.

\section{Preliminaries}
\label{sec:preliminaries}
\subsection{Large Language Models (LLMs)}
LLMs are transformer-based neural models trained on a large textual dataset. LLMs, in today's world, are used to interpret, summarize, and assist human tasks efficiently. Notable examples include OpenAI's GPT, Google's Gemini, Meta's LLaMA, and Anthropic's Claude, with each of them providing different kinds of resources and applications. LLMs can be accessed through a graphical user interface (GUI), where a user can interact with the model by querying and receiving responses directly without the need for programming. Another approach is to access LLMs through an API, which allows for integrating LLMs into software applications and workflows. 
This approach is particularly useful in software development and testing environments, where LLMs can assist in code generation, analysis, or validation tasks.
Multiple platforms, such as LLM4SoftwareTesting \cite{wang2024software}, demonstrate how LLMs can be used for testing purposes in unit testing, functional testing, system test input generation, and test oracle generation. 

Most LLM APIs provide a \emph{temperature} hyperparameter, ranging from 0 to 1, which controls the randomness of the outputs generated by the model. A higher temperature increases diversity in responses by introducing more randomness during token sampling. Meanwhile, a lower temperature generates more deterministic outputs. This parameter is useful in balancing exploration and precision depending on the task. For example, for creative or open-ended tasks, a higher temperature would suit, whereas technical or structured generation tasks generally require a lower temperature.

\subsection{Prompt Engineering}
Prompt Engineering \cite{prompt2024} is a systematic approach to design input instructions or prompts provided to an LLM to optimize its response quality and context awareness. LLMs generate responses based on the training knowledge base and input patterns or contexts. This makes it important to have a clear and structured prompt to enhance the accuracy and contextual relevance of the responses we get from the LLM. 

There are multiple prompt design techniques \cite{prompt2024}, such as \textit{zero-shot} prompting, where the model is given no examples and relies completely on its knowledge base. Alternatively, in \textit{few-shot} prompting, the model is given structured examples to get optimized results. More advanced techniques include \textit{chain-of-thought} (CoT) prompting \cite{wei2022chain}, where the model is prompted to reason through a problem before providing an output, which improves logical consistency. 

\subsection{Running Example}

To exemplify our approach, we use as a running example a simplified version of a Lubricating Oil Cooling (LOC)\footnote{https://github.com/Novia-RDI-Seafaring/fmu-library/tree/main/models/loc} system 
packed as an FMU. LOC models the heat transfer from the lubrication oil to the cooling water circuit of a marine engine unit. The LOC system has a proportional–integral (PI) controlled valve that regulates the lubrication oil temperature at a constant setpoint at the engine inlet. The controller aims to keep the lubrication oil temperature at the outlet within the specified boundary values under all operating conditions.

The Lubricating Oil Cooling system has the following input variables, 
\texttt{temperature\_cooling\_liquid\_in}, \texttt{mass\_flow\_cooling\_liquid\_in}, \texttt{engine\_load}, and \texttt{setpoint\_temperature\_oil} (held constant throughout the simulation) and provides as outputs \texttt{temperature\_oil}, \texttt{position\_valve},
\texttt{temperature\_cooling\_liquid\_out}, and \texttt{mass\_flow\_cooling\_liquid\_out}.


\section{Approach}
\label{sec:approach}

\begin{figure}[!t]
  \centering
  \includegraphics[width=\linewidth]{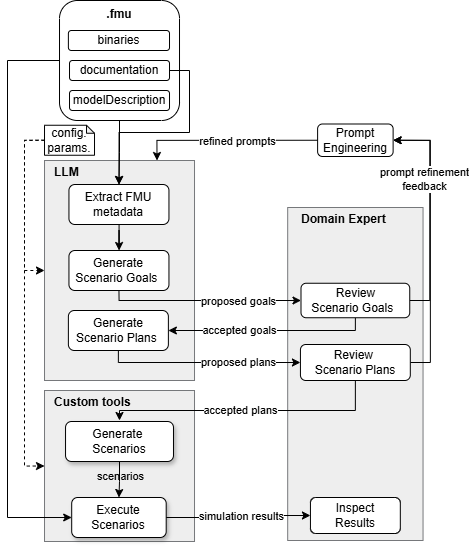}
  \caption{Overview of the Scenario Generation Approach}
  \label{fig:workflow-new}
\end{figure}

Our approach combines automatic LLM-based scenario generation with the feedback from a domain expert  (human in the loop) as shown in Figure \ref{fig:workflow-new}. The starting point is an FMI-compliant FMU and a set of customizable LLM prompts. We assume that the FMU has a well-specified \textit{modelDescription.xml} file, with complete I/O variable details, and one or several functional specification documents. We do not require access to the internal equations or solver configuration of the FMU. 

The approach follows a step-wise scenario design process divided into three abstraction levels: \emph{intent} (goals), \emph{templates} (plans), and \emph{executable time-series} (scenarios). This structure supports domain-expert review and validation at each step of the process. The instantiated simulation scenarios are then executed and validated using a simulation environment. 

Each LLM-based step is followed by post-processing to validate the generated output against the schema guidelines. This step fixes minor violations (e.g., missing null fields) and rejects the ones that are not reparable. All LLM-generated \textit{goals} and \textit{plans} are assigned deterministic identifiers (IDs) to ensure traceability of the generated scenarios throughout the process. A standalone script performs deduplication of generated goals, plans, and scenarios to ensure the test suite contains only unique entries. The script uses a hashing-based mechanism to detect exact matches, accepting a new goal only if no identical goal has been previously stored. When a duplicate is detected, the script logs the ID of the existing goal. The same hashing strategy is applied during plan and scenario generation steps to prevent redundant generations throughout the pipeline.

A domain expert reviews the generated goals to assess their feasibility and coverage of all scenarios. The generated \textit{plans} are reviewed for numerical realism and compliance with provided input-output constraints. All accepted goals and plans are then saved and hashed with IDs for scenario generation, comparisons, and deduplication in later stages.

\subsubsection{Scenario Configuration Setup} 
We define the parameters of the scenario generation lifecycle. 
It includes parameters for configuring the LLMs (\textit{model} used, \textit{temperature}) and for running the simulation (\textit{start time}, \textit{stop time}, \textit{step size}, \textit{output interval}, and \textit{allowed output tolerance}).

\subsubsection{Prompt Engineering} 
Our approach is interactive and iterative. It starts with an initial set of prompts dedicated to each phase of the workflow. The design of these prompts considers three aspects: a) it controls diversity versus determinism per phase (e.g., a higher temperature for goals to have diversity, and a lower one for plans to have determinism), b) employs context minimization techniques to improve the quality of generated results, and c) assumes that a domain expert reviews the results after each phase to enhance the overall efficiency and trustworthiness of the workflow.

Prompt templates are crafted by an LLM engineer in collaboration with a domain expert. The LLM engineer designs schema-first instructions, few-shot exemplars, and parsing constraints to ensure that LLM outputs are well-structured. The domain expert provides domain vocabulary, realistic parameter ranges, and examples that reflect realistic operating conditions. During review cycles, the domain expert also proposes prompt refinements (e.g., disallowing unsafe ramps or unrealistic input ranges), which are integrated into the next prompt versions. As such, the prompt engineering process remains both interactive and iterative.


Each prompt contains both process-specific and domain-specific instructions and follows a \emph{schema-first} structure that standardizes the information provided to the LLM. The main categories are:
\begin{itemize}[leftmargin=*]
    \item \textit{Role}: defines the professional or technical perspective the LLM should adopt (e.g., "You are an expert on FMI/FMU interfaces"). This helps to set the model's reasoning within an appropriate domain context. These defined roles are domain-specific.
    \item \textit{Task}: specifies what the model should generate (e.g., extract I/O constraints or generate scenario goals). This ensures that the response aligns with the intended task. 
    \item \textit{Success criteria}: states the expected qualities of a correct response (e.g., structured output, internal consistency, or completeness). This provides the model with an objective reference for evaluating its own output quality.
    \item \textit{Strict JSON schema}: defines the required and optional fields, data types, and allowed enumerations to ensure outputs conform to a machine-readable format.
    \item \textit{Output rule}: instructs the model to return only the JSON object, without additional prose or explanations.
    \item \textit{Few-shot exemplars}: provide short, structured examples showing correct schema usage. These domain-specific examples help the LLM to align with the expected structure.
    \item \textit{Validation constraints}: specify valid variable names, units, and ranges extracted from the FMU metadata or specification documents. These constraints are also domain-specific.
\end{itemize}

\subsubsection{Extract FMU Metadata} 
Extracts constraints for input and output variables, including names, units, causality, and allowed ranges from the \texttt{modelDescription.xml} (excerpt shown in Listing \ref{lst:model-description}) and textual specification document.

\begin{lstlisting}[language=JSON,caption={Excerpt from \texttt{modelDescription.xml} of LOC},label={lst:model-description},breaklines=true]
<ModelVariables>
  <ScalarVariable causality="input" description="Temperature of the cooling liquid at the heat exchanger inlet." name="temperature_cooling_liquid_in" valueReference="0" variability="continuous"> <Real max="100" min="0" start="0" unit="degC"/>
  </ScalarVariable>
  <ScalarVariable causality="output" description="Temperature of the cooling liquid at the outlet of the heat exchanger." initial="calculated" name="temperature_cooling_liquid_out" valueReference="4" variability="continuous"> <Real max="100" min="0" unit="degC"/>
  </ScalarVariable>
</ModelVariables>
\end{lstlisting}

Listing \ref{lst:prompt-constraints} provides the prompt used to extract the I/O constraints from FMU metadata, while Listing \ref{lst:extracted-constraints} shows the provided response with all I/O constraints.




\begin{lstlisting}[caption={I/O Constraints Extraction Prompt}, label={lst:prompt-constraints}]
1. Role:
You are an expert on FMI/FMU interfaces.

2. Task:
From the following documentation, extract all *input* and *output* variables with their numeric ranges for the system {system_name}.fmu.

3. Rules:
- Always include "name", "min", "max", and "unit" (if mentioned).
- If no explicit min/max is given, return null for that bound.
- Sometimes Range is given in **Min-Max (Initial Value/Calculated)** format, so check that carefully.
- Units must be preserved exactly if present (e.g., kg/s).

4. JSON Schema:
- Respond with JSON only in this exact format:
{{ "inputs": [
    {{"name": "...", "min": ..., "max": ..., "unit": "..."}}
  ],
  "outputs": [
    {{"name": "...", "min": ..., "max": ..., "unit": "..."}}
  ] }}

5. Context Document:
{merged_doc}
\end{lstlisting}

\begin{lstlisting}[caption={Extracted I/O Constraints for LOC FMU},label={lst:extracted-constraints}]
{"inputs": [
    {"name": "temperature_cooling_liquid_in", "min": 0, "max": 100, "unit": "degC"},
    {"name": "mass_flow_cooling_liquid_in", "min": 0, "max": 50, "unit": "kg/s"},
    {"name": "setpoint_temperature_oil", "min": 30, "max": 90, "unit": "degC"},
    {"name": "engine_load", "min": 0,"max": 1, "unit": ""}],
  "outputs": [
    {"name": "temperature_cooling_liquid_out", "min": 0, "max": 100, "unit": "degC"},
    {"name": "mass_flow_cooling_liquid_out", "min": 0, "max": 50, "unit": "kg/s"},
    {"name": "temperature_oil", "min": 0, "max": 100, "unit": "degC"},
    {"name": "position_valve", "min": 0, "max": 1, "unit": ""} ] }
\end{lstlisting}

Although extraction is performed by the LLM, the schema-first prompt strictly constrains the LLM-generated response to data given in the provided input specifications. In our runs, repeated executions generated the same set of variables and preserved names/units, confirming that only the FMU’s actual inputs and outputs are returned. This allows us to proceed without a domain expert review specifically for this step.

\subsubsection{Generate Scenario Goals} Prompts the LLM to generate a diverse set of \textit{scenario goals} in the form of Gherkin's \emph{Given-When-Then (GWT)} patterns. Gherkin is a 
language that uses a simple, easy-to-read structure for writing software test cases \cite{wynne2012cucumber}. 
GWT patterns provide domain experts with a user-friendly option to review generated goals in a human-readable format. Our previous work on testing dynamic simulation models using metamorphic testing \cite{sudheerbabu2025validation} showed that such a format is easier to adopt by the domain expert when formulating tests. 
Furthermore, empirical evaluations showed that LLMs provide better results with a structured given conditions compared to unformatted text. 
In our approach, We use GWT patterns to describe a scenario's preconditions and initial context (\textit{given}), the action that triggers a behavioural change (\textit{when}), and the expected outcome or result (\textit{then}). 



However, one challenge that remains is the automatic creation of the expected output (aka. the \textit{test oracle problem}) of a dynamic simulation. This is particularly difficult since the value of each output has to be accurately predicted at each time step of the simulation, and it is heavily influenced by the internal structure of the model and domain knowledge. 



To address this problem, we define generic oracles 
which describe expected changes or trends in the outputs instead of concrete values. 
These will be used later on to compare the expected and actual outputs of the simulation. The oracles and their semantics are defined as follows: Let $y(t)$ be the output time series sampled on $t \in [t_{\text{start\_time}}, t_{\text{stop\_time}}]$.
\begin{itemize}
  \item \texttt{bounded(low, high)}: PASS if the values stay between low and high bounds during the while simulation period.
  \item \texttt{crosses\_above(threshold,by\_time)}: PASS if the value goes above the threshold before the defined by\_time.
  \item \texttt{crosses\_below(threshold, by\_time)}: PASS if the value passes below the threshold before the defined by\_time.
  \item \texttt{monotonic\_increase(eps)}: PASS if, for all successive samples, the value is greater than the previous value (tolerant to small numerical noise, eps).
  \item \texttt{monotonic\_decrease(eps)}: PASS if, for all successive samples, the value is less than the previous value (tolerant to small numerical noise, eps).
  \item \texttt{settles\_to(target,tol,within)}: PASS if the value settles to the target variable within the defined time limit, with an allowed tolerance, tol.
\end{itemize}

Listing \ref{lst:prompt-goals} shows the prompt used for goal generation, while Listing \ref{lst:example-goal} shows the generated scenario which includes a \textit{scenario rationale}, \texttt{scenario\_count}, and \texttt{scenario\_count\_rationale}, explaining why this scenario was provided, what a suitable scenario count might be, and the rationale for that count.



\begin{lstlisting}[caption={Goals Generation Prompt},label={lst:prompt-goals}]
1. Role:
You are planning **Scenario Goals** for the FMI-based system "{system_name}.fmu".

2. Context Document
System metadata/description:
---
{metadata_text}
---

3. I/O Constraints
I/O constraints (use ONLY variable names that appear here):
{constraints_json}

4. Task
Your task:
- Generate a **concise, distinct, and extensive** set of behaviour-focused goals in **Given-When-Then** form.
- Each goal MUST include: id, pattern, given, when, then, goal_rationale, target_count, target_count_rationale.
- 'pattern' MUST be exactly "Given-When-Then".
- 'target_count' is how many **distinct test plans** will be generated later for this goal.

5. Rules
Hard constraints:
- Use ONLY variable names that actually appear in the constraints or metadata.
- If an input contains 'setpoint' in its name, KEEP IT CONSTANT in **Given** and DO NOT CHANGE IT in **When**.
- Do NOT include numbers, time-series arrays, setpoint values, thresholds, times, or units anywhere.
- Do NOT invent variables or constants.

6. General Guidelines
Rules for GWT:
- **Given**: qualitative initial conditions for relevant *INPUT* variables (setpoints fixed).
- **When**: qualitative disturbance on one or more non-setpoint *INPUT* variables (step/ramp).
- **Then**: an **ARRAY**; follow these rules strictly:

i) THEN WRITING RULES:
- One THEN item = exactly one **OUTPUT** variable.
- Express the strongest single behaviour (priority): 
    1) settles_to <SETPOINT_INPUT> (and stays there for the remaining simulation time)
    2) crosses_above baseline
    3) increases monotonically / decreases monotonically
    4) remains within its valid range
- If an OUTPUT maps to a setpoint INPUT (e.g., temperature_oil - setpoint_temperature_oil), prefer:
    "temperature_oil settles to setpoint_temperature_oil and remains within its valid range thereafter."
- NO numbers, thresholds, times, or units. 
- NO vague verbs: "responds", "changes", "adjusts", "regulates".
- Do not mention more than one OUTPUT in a single THEN item.

ii) Coverage:
- For each INPUT (excluding setpoints), create all possible goals that cover:
  -- Increase and decrease directions (where meaningful).
  -- At least one step and one ramp disturbance (when target_count > 1, use these as distinct plans).
- Include at least one multi-disturbance goal (sequential changes on >1 INPUT) without altering setpoints.
- Never change any setpoint input in Given/When.

iii) Additional notes:
- Prioritize information from sections like "Behaviour, constraints and realistic rates" and "Model description, I/O and Parameters".
- Use exact variable names; do not invent signals.

iv) Coverage Instructions:
Limit your goals to selected test type(s) {types_str}

7. Strict Schema
STRICT SCHEMA (no extra fields; use exact field names):
{  "goals": [
    {
      "id": "G001",
      "pattern": "Given-When-Then",
      "given": "Describe initial INPUT state; use exact names; keep setpoints constant; no numbers.",
      "when": "Describe the INPUT disturbance (step or ramp) on non-setpoint input(s); no numbers.",
      "then": [
        "Describe the change in Output(s) based on the change triggered by the 'when' condition."
      ],
      "goal_rationale": "Short reason.",
      "target_count": 1,
      "target_count_rationale": "Short reason."
    }   ] }
\end{lstlisting}

\begin{lstlisting}[language=json,caption={Example - LLM-generated Goal (G001)},label={lst:example-goal}]
{    "id": "G001",
    "pattern": "Given-When-Then",
    "given": "temperature_cooling_liquid_in, mass_flow_cooling_liquid_in, engine_load are at nominal values; setpoint_temperature_oil is constant.",
    "when": "engine_load is increased with a step change.",
    "then": [
        "temperature_oil settles to setpoint_temperature_oil and stays there for the remaining simulation time.",
        "temperature_cooling_liquid_out increases monotonically.",
        "position_valve decreases monotonically.",
        "mass_flow_cooling_liquid_out remains within its valid range."    ],
    "goal_rationale": "Verifies closed-loop regulation and output responses to a sudden increase in engine_load.",
    "target_count": 1,
    "target_count_rationale": "A single step increase covers primary regulation behavior."}
\end{lstlisting}

\subsubsection{Generate Scenario Plans}
The approved scenario goals are converted by the LLM into \textit{scenario plans} with parameter space (\texttt{param\_space}) for input variables and assertions for the outputs. For each accepted scenario goal, the LLM provides one or more scenario \textit{plans} with input patterns (\textit{constant/step/ramp}), magnitudes, change times, and an assertion window. 
The complete prompt to the LLM for plan generation is shown in 
Listing \ref{lst:prompt-plans}, while the corresponding LLM-generated plan for G001 is shown in Listing \ref{lst:example-plan}.


\begin{lstlisting}[caption={Plans Generation Prompt},label={lst:prompt-plans},breaklines=true]
1. Role:
You are a Test Generation expert producing **Scenario Plans** for the FMI system "{system_name}.fmu".

2. Context
- Simulation time window (seconds): start={sim_start}, stop={sim_stop}. Do not use any time outside this window.
- System I/O constraints (use exact variable names and stay within these numeric bounds):
{constraints_json}

3. Goals (use these exactly; do NOT invent new goals):
{json.dumps(goals_brief, ensure_ascii=False, indent=2)}

4. Schema
SCHEMA YOU MUST FOLLOW (no extra fields):
{{"plans": [
    {{"goal_id": "Gxxx",
      "type": "positive" | "boundary",
      "param_space": {{
        "<INPUT_VAR>": {{ "pattern": "constant", "value": <num | [min,max]> }} |
                        {{ "pattern": "step", "from": <num>, "to": <num | [min,max]>, "at": <num | [t_min,t_max]> }} |
                        {{ "pattern": "ramp", "start": <num>, "end": <num | [min,max]>, "duration": <num | [min,max]> }} }},
      "assertions": [
        {{ "kind": "bounded", "var": "<OUTPUT_VAR>", "low": <num>, "high": <num>, "from_timestep"?: <t>, "to_timestep"?: <t> }} |
        {{ "kind": "crosses_above", "var": "<OUTPUT_VAR>", "threshold": <num>, "by_time": <t> }} |
        {{ "kind": "monotonic_increase", "var": "<OUTPUT_VAR>", "from_timestep"?: <t>, "to_timestep"?: <t>, "eps"?: <num> }} |
        {{ "kind": "monotonic_decrease", "var": "<OUTPUT_VAR>", "from_timestep"?: <t>, "to_timestep"?: <t>, "eps"?: <num> }} |
        {{ "kind": "settles_to", "var": "<OUTPUT_VAR>", "target": <num>, "tol": <num>, "within": <t> }} |
        {{ "kind": "settles_to", "var": "<OUTPUT_VAR>", "target_var": "<INPUT_SETPOINT_VAR>", "tol": <num>, "within": <t> }} ] }} ] }}

5. Requirements
A) Mapping each THEN item with exactly one assertion for that OUTPUT (no duplicates for the same var):
   - If THEN mentions **settles to <SETPOINT_INPUT>**, use ONLY `settles_to` with `target_var="<SETPOINT_INPUT>"`.
     Do NOT also add monotonic or bounded assertions for that same output.
     Interpret `settles_to` as: output enters ±tol of the target by `within` and remains within that band until sim_stop.
   - Otherwise, choose exactly one of: `crosses_above`, `monotonic_increase`, `monotonic_decrease`, or `bounded`.

B) Setpoint discipline:
   - Never change any setpoint input in `param_space` (setpoints may appear only as `target_var` in `settles_to`).
   - Use only non-setpoint inputs as drivers.

C) No hallucinations:
   - Use only variable names from the constraints. If a name is absent, do not use it.

D) Numeric choices and timing so settling can succeed:
   - Place main step/ramp start early: choose 'step.at' or begin of 'ramp' in approx 10-25% of the window
   - Choose 'settles_to.within' around approx 60-85% of the window (<= {sim_stop}), leaving enough time to settle and remain.
   - Keep all times within [{sim_start},{sim_stop}]. For ramps, 'duration' belongs to [0, stop-start].
   - Keep numbers consistent with the constraints; ≤3 decimals.

E) Plan distinctness:
   - For goals with 'target_count' > 1, vary at least one of: driver input, direction (increase/decrease),
     pattern (step/ramp), start time, or amplitude.
   - Avoid duplicates within this response{avoid_hint}

F) Sanity:
   - Inputs should plausibly drive the chosen output behaviour.
   - Use `bounded` when it is the sole THEN behaviour.

6. Output Rules
- Return ONLY the JSON object described above (no prose, no comments).
- Avoid duplicates of these already-accepted plans (per goal): {avoid_text}
\end{lstlisting}

\begin{lstlisting}[language=json,caption={Example - LLM-generated Plan for G001 (P001)},label={lst:example-plan}]
{    "id": "G001-P001",
      "param_space": {
        "temperature_cooling_liquid_in": {"pattern": "constant", "value":[50.0]},
        "mass_flow_cooling_liquid_in": {"pattern": "constant", "value":[25.0]},
        "setpoint_temperature_oil": {"pattern": "constant", "value":[70.0]},
        "engine_load": {"pattern": "step", "from": 0.5, "to":[0.9], "at":[150.0]}      },
      "assertions": [
        {"kind": "settles_to", "var": "temperature_oil", "target_var": "setpoint_temperature_oil", "tol": 1.0, "within": 700.0},
        {"kind": "monotonic_increasing", "var": "temperature_cooling_liquid_out", "from_timestep": 150.0, "to_timestep": 999.0, "eps": 0.05},
        {"kind": "monotonic_decreasing", "var": "position_valve", "from_timestep": 150.0, "to_timestep": 999.0, "eps": 0.01},
        {"kind": "bounded", "var": "mass_flow_cooling_liquid_out", "low": 0.0, "high": 50.0}      ]}
\end{lstlisting}

\subsubsection{Domain Expert Validation and Feedback}


The domain expert performs a two-phase review process covering both scenario \texttt{goals} and \texttt{plans}. In the first phase, the domain expert verifies that the LLM-generated \texttt{goals} are consistent with system dynamics and discards any \texttt{goals} that are unrealistic or behaviourally incorrect. In the second phase, the domain expert reviews the generated plans, ensuring that safety constraints, parameter ranges, and input–output relationships are correct. Accepted \texttt{goals} and \texttt{plans} are then finalized and passed on to the next phase.

\subsubsection{Generate Scenarios} 


In this phase, scenario plans are converted into \textit{simulation scenarios} by assigning concrete values to the input and the output variables. We use the Latin Hypercube Sampling (LHS) method \cite{shields2016generalization} with a fixed random seed to ensure deterministic generation under the same plans and configuration. 
Each scenario is assigned a unique ID, G\textit{xxx}-P\textit{yyy}-T\textit{zzz}, along with an input hash used for both within-run and cross-run deduplication (removing redundant plans) to avoid redundant executions. With a fixed random seed, LHS remains deterministic, ensuring that the same plan and constraints produce the same sampled magnitudes and times. This is crucial for deduplication as we compute an input-hash for all instantiated scenarios. Scenarios with matching input hashes are skipped within and across runs. In our setup, we use \emph{seed}~=~42, and generate one instantiation per plan. A complete scenario is presented in Listing \ref{lst:example-scenario}.

\begin{lstlisting}[language=json,caption={Scenario for G001-P001 (T001)},label={lst:example-scenario}]
{    "test_id": "G001-P001-T001",
    "inputs": {
        "engine_load": [...],
        "setpoint_temperature_oil": [...],
        "temperature_cooling_liquid_in": [...],
        "mass_flow_cooling_liquid_in": [...]    },
    "assertions": [
        {"kind": "settles_to", "var": "temperature_oil", "target_var": "setpoint_temperature_oil", "tol": 1.0, "within": 700.0},
        {"kind": "monotonic_increasing", "var": "temperature_cooling_liquid_out", "from_timestep": 150.0, "to_timestep": 999.0, "eps": 0.05},
        {"kind": "monotonic_decreasing", "var": "position_valve", "from_timestep": 150.0, "to_timestep": 999.0, "eps": 0.01},
        {"kind": "bounded", "var": "mass_flow_cooling_liquid_out", "low": 0.0, "high": 50.0}    ] }
\end{lstlisting}

\subsubsection{Execute Scenarios} 
Each simulation scenario is executed within the simulation environment using the \textit{simulation time} and \textit{step size} defined in the configuration parameters. The outputs are recorded at each time step, and all simulation settings (including step, solver defaults, and tolerances) are logged for traceability. 

For each scenario, the assertions defined in the plan are evaluated over their defined time windows. 
After execution, Boolean verdicts are assigned to each assertion and then aggregated at the scenario level.
A scenario is considered passed only if all assertions defined in the plan evaluate to true; otherwise, it is marked as failed.





\subsubsection{Inspect Results} The results of the simulation are logged and provided to the domain expert for further analysis of the failed scenarios. 

\section{Evaluation}
\label{sec:validation}

\subsubsection{Experimental setup}

We use OpenAI's GPT model, \textit{gpt-4.1}, via an API. We set a higher temperature (0.7) for goal generation to have diversity in the generated goals and a lower temperature (0.2) for the plan generation phase to increase determinism. We use the \textit{FMPy} Python library \cite{FMPy} to simulate the FMU. 




\subsubsection{Results}
We executed the toolchain 6 times without human intervention and collected the results shown in Table~\ref{tab:gen-stats}. Some of the generated goals, plans, and scenarios produced in each phase were incorrect and had to be edited and/or discarded. The earlier in the process the erroneous artefacts were detected, the more adequate the scenarios were. By the nature of the LLMs, the output they produce at each step against a given prompt differs from one session to another, even when the temperature parameter is set to a minimum. For instance, in some sessions, 7 goals are produced while in others, 8. Without domain expert review and deduplication, some of the inconsistent goals will be propagated to later phases and will produce incorrect scenarios, which in our experiments resulted in 47\% average accuracy. After including the domain expert, the accuracy increased to nearly 100\%.  
Therefore, it is crucial that the output of the LLM is considered advisory, and it should be continuously supervised by the human user.

\begin{table}[h]
\centering
\begin{tabular}{|r|r|r|r|r|}\hline
\textbf{Goals} & \textbf{Plans} & \textbf{Valid Scenarios} & \textbf{Invalid Scenarios} & \textbf{Accuracy} \\\hline
7 & 7 & 4 & 3 & 0.57\\\hline
7 & 7 & 3 & 4 & 0.43\\\hline
7 & 13 & 5 & 8 & 0.38\\\hline
8 & 9 & 7 & 2 & 0.78\\\hline
7 & 10 & 3 & 7 & 0.30\\\hline
8 & 8 & 3 & 5 & 0.38\\ \hline
\end{tabular}
\caption{Phase-wise Generation Statistics}
\label{tab:gen-stats}
\end{table}

Because the LLM sampling process is inherently stochastic, the number and content of generated goals and plans may vary between runs, even for identical prompts and temperature settings. During the domain expert review, further changes are made in the generations, so the final counts are not guaranteed to be stable. However, the scenario generation step is fully deterministic. Given a fixed set of accepted plans, simulation configuration, and seed value, LHS produces the same parameter values and the same number and content of scenarios in repeated runs.


Following a manual approach, the domain expert created 6 goals and valid scenarios (see \cite{sudheerbabu2025validation}) from the specification, which were largely covered by the generated goals and plans, but with different values for the scenarios. However, our approach generated additional scenarios compared to the manual approach, which were different both in terms of plans and goals. 

 


\subsubsection{Scenario Adequacy}
\begin{figure}
     \centering
     \includegraphics[width=0.6\linewidth]{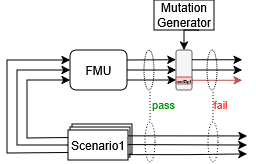}
     \caption{Mutation Generation Approach}
     \label{fig:placeholder}
 \end{figure}

To evaluate the quality of the generated scenarios, we apply mutation analysis. Since we do not have access to the internal specification of the FMU, we imitate possible design and implementation mistakes by applying systematic changes (\textit{mutations)} in the outputs (see Figure \ref{fig:placeholder}). These mutations are defined based on a set of \textit{mutation operators} that are systematically applied, one at a time, to each output; each application of one operator to an output will result in a slightly incorrect version (\textit{mutant}) of the FMU.  We use the following mutation operators:

\begin{itemize}
    \item \textit{Mirror Mutation} : Replaces an output value at the middle point of the boundary interval with its mirror value.
    \item \textit{Random Uniform Mutation}: Replace values of a selected time-step range 
    with a uniform random value for a chosen output signal time-series.	
    \item \textit{Crossover Mutation}: Select two output time series for a given output signal and pick a specific crossover site and 
    interchange the values of the time series signals after the crossover site. 
    \item \textit{Polynomial Mutation}: Given a variable x in a time series in the range ($x_{min}$, $x_{max}$), we slightly change it by a random amount determined by a polynomial distribution.		     
\end{itemize}

The process starts with the execution of the generated scenarios on the original version of the FMU.  If there are failing scenarios, they are either removed or the FMU is fixed. To save on the simulation time, we apply the mutation operators directly on the recorded output time series for each output that is evaluated in a scenario evaluating (this will reduce the number of mutants generated). Each application of a mutation operand to an output time series will result into a new mutant. We evaluate the expected output values from the scenario with the ones in the mutant and assign a pass/fail verdict. Finally,  we calculate the adequacy as a \textit{mutation score}, with values between 0 and 1, as the ratio between the number of mutants on which some scenarios failed (killed mutants) and the total number of mutants created. 


We applied this approach to the set of manually created scenarios and to a set of 6 scenarios generated with our approach.  The mutation generation process resulted in 70 mutants of LOC. The manually generated scenarios killed 48 mutants, resulting in a mutation score of 0.685, whereas the scenarios generated with our approach killed 47 mutants, resulting in a mutation score of 0.67. This shows that the generated scenarios have the same fault-detection capability compared to the manually created ones. This can be explained by the fact that the domain expert guardrails the LLM-generated output at different phases of the process. However, further investigation into the effectiveness of the mutation operators used is needed, with more complex FMUs. 
  

\textit{Runtime and token limits:} For our LOC example, typical runtime is \(\approx 5\text{–}7\) s for constraints extraction, \(\approx 15\text{–}25\) s for scenario goals, and \(\approx 25\text{–}30\) for scenario plans. End-to-end generation (constraints, goals, and plans for a batch of 8-10 scenarios) remained under one minute per run in our case. 

\section{Conclusions and future work}

The results highlight several key insights about the proposed approach. One of the most significant findings is the importance of \textbf{domain expert review}. While LLMs can autonomously extract constraints, generate goals, and plans with reasonable structure, the absence of expert validation leads to semantically weak or inconsistent scenarios. In particular, some goals failed to capture the intended system behaviour or relied on unrealistic oracles. The domain expert review ensures that the generated goals and plans align with actual system behaviour. This also ensures that unsafe or infeasible scenarios are discarded early. This human-in-the-loop step effectively bridges the gap between LLM generation and domain realism.

Another key finding is the impact of \textbf{prompt refinements}. Early iterations using generic prompts produced incomplete or inconsistent JSON structures. Introducing schema-first templates, few-shot examples, and detailed operator semantics (such as the correct interpretation of \texttt{settles\_to}) improved structural correctness. Moreover, specifying the simulation time window explicitly in the prompts helped the LLM generate more realistic assertion thresholds. These refinements show how iterative prompt engineering directly contributes to improved scenario quality.

The results also highlight the effectiveness of the \textbf{temperature-split strategy}. Using a higher temperature during goal generation provided diversity, allowing the model to propose a broader set of behaviourally distinct scenario goals. Meanwhile, a lower temperature in the plan generation phase ensured determinism and precision in parameter values and time windows. This balance between diversity and determinism helped to generate a comprehensive and distinct set of scenarios that test the system behaviour thoroughly.

Another contribution of the approach is the \textbf{deduplication process}. Since LLMs may regenerate semantically identical goals and plans across runs, our deduplication process (both within-run and cross-run) helped to maintain efficiency. With this, the approach ensures behavioural coverage while reducing simulation time and redundant evaluations.

However, there are still several limitations that affect the robustness and scalability of the approach. First, \textbf{non-determinism} is inherent to LLMs, meaning that slight variations in temperature, sampling, or context order can lead to different results even under identical conditions. While having a strict schema reduces variability, complete reproducibility across runs cannot be guaranteed. In addition, \textbf{hallucinations} create results where the model assigns unrealistic parameter ranges or incorrect formatting. Domain expert review handles these cases, but it cannot fully eliminate them.

Finally, there is the challenge of \textbf{trustworthiness in oracle generation and verdict assignment}. Since the LLM generates qualitative oracles (e.g., “settles to” or “bounded”) rather than numerical ground truth, their validity depends on how well these operators map to the system’s actual dynamics. If thresholds or time windows are set too tightly, correct system behaviours may be incorrectly flagged as failures. This limitation highlights the need for adaptive or data-driven oracle calibration, which also remains an open research problem.

\label{sec:conclusions}
Future work includes extending the approach to co-simulation FMUs and multi-FMU setups, and adding an active learning loop that re-prompts the LLM from failures.


\bibliographystyle{IEEEtran}
\bibliography{mybibfile}



\end{document}